\documentclass[12pt]{article}
\usepackage{graphicx,amsmath,amssymb}
\newcommand{\cinst}[2]{$^{\mathrm{#1)}}$~#2\par}
\newcommand{\crefi}[1]{$^{\mathrm{#1)}}$}
\newcommand{\HRule}{\rule{0.4\linewidth}{0.3mm}}
\newcommand{\jpsi}{J/$\psi$}
\newcommand{\xf}{$x_{\mathrm{F}}$}
\newcommand{\pt}{$p_{\mathrm{T}}$}
\newcommand\jhep[3]{{{J.\ High\ Energy\ Phys.\ }{\bf #1} (#2) #3}}
\newcommand\myprd[3] {{{Phys.\ Rev.\ }{\bf D #1} (#2) #3}}
\newcommand\myprl[3] {{{Phys.\ Rev.\ Lett.\ }{\bf #1} (#2) #3}}

\begin{document}

\begingroup
\thispagestyle{empty} \baselineskip=14pt
\parskip 0pt plus 5pt

\begin{center}
{\large EUROPEAN LABORATORY FOR PARTICLE PHYSICS}
\end{center}

\bigskip
\begin{flushright}
CERN--PH--EP\,/\,2008--024\\
December 10, 2008
\end{flushright}

\bigskip
\begin{center}
{\Large\bf \boldmath \jpsi\ polarization\\[5mm] from fixed-target to collider
  energies}

\bigskip\bigskip

Pietro Faccioli\crefi{1},
Carlos Louren\c{c}o\crefi{2},
Jo\~ao Seixas\crefi{1,3}
and Hermine K.\ W\"ohri\crefi{1}

\bigskip\bigskip\bigskip
\textbf{Abstract}

\end{center}

\begingroup
\leftskip=0.4cm \rightskip=0.4cm
\parindent=0.pt

The determination of the magnitude and ``sign'' of the \jpsi\
polarization crucially depends on the reference frame used in the
analysis of the data and a full understanding of the polarization
phenomenon requires measurements reported in \emph{two} ``orthogonal''
frames, such as the Collins-Soper and helicity frames.  Moreover, the
azimuthal anisotropy can be, in certain frames, as significant as the
polar one.
The seemingly contradictory \jpsi\ polarization results reported by
E866, HERA-B and CDF can be consistently described assuming that the
most suitable axis for the measurement is along the direction of the
relative motion of the colliding partons, and that directly produced
\jpsi's are longitudinally polarized at low momentum and transversely
polarized at high momentum.
We make specific predictions that can be tested on existing CDF data
and by LHC measurements, which should show a full transverse
polarization for direct \jpsi\ mesons of $p_\mathrm{T} > 25$~GeV$/c$.

\endgroup

\begin{center}
\emph{Submitted to Phys. Rev. Lett.}
\end{center}

\vfill
\begin{flushleft}
\HRule\\

\cinst{1} {Laborat\'orio de Instrumenta\c{c}\~ao e F\'{\i}sica Experimental de
  Part\'{\i}culas (LIP),\\ ~~~Lisbon, Portugal} 
\cinst{2} {CERN, Geneva, Switzerland}
\cinst{3} {Instituto Superior T\'ecnico (IST) and Centro de  F\'{\i}sica Te\'orica de
  Part\'{\i}culas\\ ~~~(CFTP), Lisbon, Portugal}
\end{flushleft}
\endgroup

\newpage
\sloppy

\begin{figure}[h]
\centering
\includegraphics[width=0.65\linewidth]{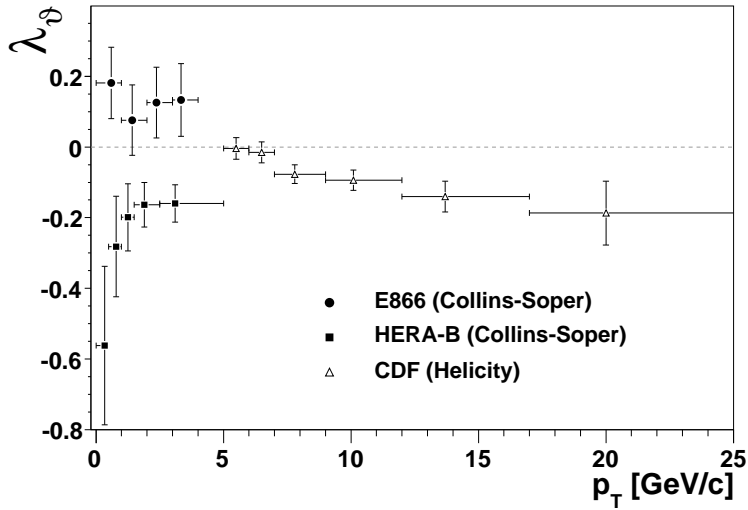}
\vglue-2mm
\caption{$\lambda_{\vartheta}$ versus $p_\mathrm{T}$, as reported by
  E866, HERA-B and CDF (statistical and systematic errors added in
  quadrature).}
\label{fig:all_data_pt}
\end{figure}

The existing measurements of \jpsi\ polarization in hadronic
collisions represent one of the most difficult challenges currently
faced by models of quarkonium production (see, for example,
Refs.~\cite{bib:Theory,bib:cdf} and references therein).  The often
emphasized disagreement between experiment and theory is, however,
only one aspect of the problem.  The experimental knowledge itself
looks contradictory when different polarization measurements are
compared, in terms of ``sign'', magnitude and kinematic dependence, as
illustrated in Fig.~\ref{fig:all_data_pt}, which shows the data
reported by CDF~\cite{bib:cdf}, HERA-B~\cite{bib:herab} and
E866~\cite{bib:e866}.

Besides the obvious interest of understanding the mechanisms of
quarkonium polarization, having a clear (data driven) description of
the polarization measurements is also important to evaluate detector
specific corrections needed to extract physics results from the data.
Production cross sections, for instance, might significantly depend on
the polarization scenario used in the calculation of acceptance
corrections.
The polarization measurements are undeniably complex and involve
difficult experimental problems.  There is, however, an additional
cause for the blurred picture emerging from the comparison of the
existing measurements: different experiments have often chosen
different polarization frames to perform their analyses.
The influence of such choices on the measured
angular distribution of the decay leptons is generally underestimated.  In
fact, different analyses of \emph{the same} two-body angular decay
distribution may give qualitatively and quantitatively different
results depending on the definition of the polarization frame.

Several polarization frame definitions have been used in the past.
In the helicity frame the polar axis coincides with the flight
direction of the \jpsi\ in the centre-of-mass frame of the colliding
hadrons.  A very different approach is implicit in
the definition of the Collins-Soper~\cite{bib:coll_sop} frame, where
the polar axis reflects, on average, the direction of the relative
velocity of the colliding partons, the approximation being especially
good if we can neglect the smearing effect due to the parton intrinsic
transverse momentum.
We denote by $\vartheta$ the angle between the direction of the
positive lepton and the chosen polar axis, and by $\varphi$ the
azimuthal angle, measured with respect to the plane
formed by the momenta of the colliding hadrons in the \jpsi\ rest
frame (the ``production plane'').
The angular decay distribution, symmetric with respect to the
production plane and invariant under parity
transformation~\cite{bib:coll_sop, bib:gott_jack}, 
is usually defined as:
%
\begin{eqnarray}
\frac{{\rm d}N}{{\rm d}(\cos\vartheta) \, {\rm d}\varphi} \; \propto 
  1 + \lambda_\vartheta
  \, \cos^2\vartheta + \lambda_{\vartheta \varphi} \, \sin 2 \vartheta
  \, \cos \varphi \\ \nonumber
  + \, \lambda_\varphi \, \sin^2\vartheta \, \cos 2
  \varphi \quad . \label{eq:angdistr} 
\end{eqnarray}
If the \jpsi\ is observed in a given
kinematic configuration, any two polarization frames differ only by a
rotation around the axis perpendicular to the production plane (the
``$y$ axis'').  The functional dependence of the decay distribution on
the angles $\vartheta$ and $\varphi$ is invariant with respect to such
a rotation, but the numerical values of $\lambda_{\vartheta}$,
$\lambda_{\vartheta \varphi}$ and $\lambda_{\varphi}$ change in a
correlated way.
In particular, a rotation by the angle $\delta = 1/2 \arctan [ 2 \,
\lambda _{\vartheta \varphi} / ( \lambda _\varphi - \lambda _\vartheta
) ]$ (or $45^\circ$ when $\lambda_\varphi = \lambda_\vartheta$) leads
to a frame where $\lambda_{\vartheta \varphi}$ is zero, i.e., a frame
with axes along the principal symmetry axes of the polarized angular
distribution.  The experimental determination of $\lambda_{\vartheta
  \varphi}$ can, therefore, provide a criterion for the choice of a
particularly convenient reference frame for the description of the
angular distribution.

While all three coefficients provide interesting and independent
information, most available measurements of J$/\psi$ polarization are
restricted to $\lambda_{\vartheta}$.  This limits the possible
interpretations of the results and forces us to rely on
model-dependent assumptions when comparing results obtained by
experiments using different reference frames.
Even the seemingly simple classification of ``transverse'' or
``longitudinal'' polarization\,\footnote{Following a common (albeit
  misleading) practice, the polarization is defined as transverse
  (longitudinal) when $\lambda_{\vartheta}>0$
  ($\lambda_{\vartheta}<0$).}
is, in fact, dependent on the reference
frame.  This is particularly evident when the decaying particle is
produced with small longitudinal momentum, the Collins-Soper (CS) and
helicity (H) polar axes becoming perpendicular to each other.
In this case (assuming $\lambda_{\varphi} = \lambda_{\vartheta
  \varphi} = 0$, for simplicity), if in one frame we measure a value
$\lambda_{\vartheta}$, the value measured in the second frame is
smaller and of opposite sign, $\lambda_\vartheta^\prime = -
\lambda_\vartheta / ( 2 + \lambda_\vartheta)$, while an azimuthal
anisotropy appears, $\lambda_\varphi^\prime = \lambda_\vartheta / ( 2
+ \lambda_\vartheta)$.

There is a further reason for performing the experimental analyses in more
than one reference frame.  The \jpsi\ acquires its polarization with
respect to a ``natural'' polarization axis which is, a priori, unknown
and not necessarily definable event-by-event in terms of observable
quantities.  In practice, a fine-grained scan of the multidimensional
phase-space of the \jpsi\ production process is not possible, due to the
limited sample of collected events, which forces the decay distribution to
be measured as an average over a wide spectrum of kinematic
configurations.
This means that the orientation of the polar axis of the chosen frame
with respect to the ``natural axis'' changes from event to event,
depending on the momentum of the produced \jpsi.  The resulting
superposition of many distributions, equal in shape but randomly
rotated with respect to one another, is ``smeared'' into a more
spherically symmetric shape.  As a consequence, the measured absolute
values of $\lambda_{\vartheta}$ and $\lambda_{\varphi}$ are smaller
than what would be measured in a fixed kinematic configuration and in
the ``natural frame''.
Therefore, independently of any prior theoretical expectation, the
frame closest to the natural frame is the one providing the smallest
$\delta$ angle \emph{and} the most significant
$|\lambda_{\vartheta}|$.

The HERA-B experiment recently reported the three coefficients
determining the \jpsi\ decay angular distribution, in three
reference frames~\cite{bib:herab}, providing a clear picture of how the
shape of the distribution changes from frame to frame.
Before discussing kinematical dependences, we start by considering the
values integrated in the phase space window covered by HERA-B:
in the CS frame,
$\lambda_{\vartheta} = -0.31 \pm 0.05$ and $\lambda_{\varphi} = -0.02
\pm 0.02$; in the H frame, $\lambda_{\vartheta} = -0.11 \pm
0.05$ and $\lambda_{\varphi} = -0.07 \pm 0.02$ (statistical and
systematic errors added in quadrature).
Furthermore, $\delta$ has a much larger error in the H frame
($10^\circ\pm20^\circ$) than in the CS frame ($3^\circ\pm3^\circ$),
reflecting the poorer precision with which the ``tilt'' of a more
spherically symmetric shape can be determined.
With the largest $|\lambda_{\vartheta}|$ and a $\lambda_{\varphi}$
compatible with zero, the CS frame is shown by the HERA-B
measurements to provide a simpler angular distribution than the
H frame.

\begin{figure}[h]
\centering
\includegraphics[width=\textwidth]{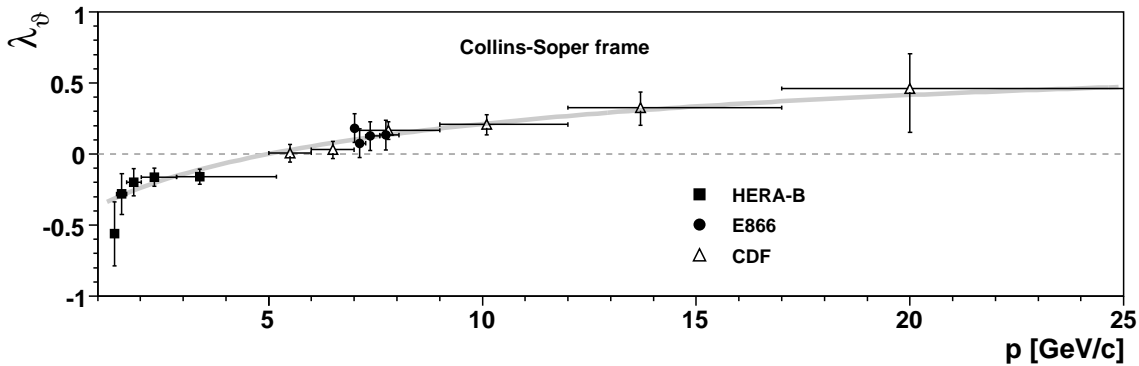}
\vglue-2mm
\caption{$\lambda_{\vartheta}$ as a function of $p$, from E866, HERA-B
  and CDF data (statistical and systematic errors added in
  quadrature).}
\label{fig:all_data_p}
\end{figure}

We now address the kinematical dependence of the \jpsi\ polarization.
Figure~\ref{fig:all_data_pt} shows that, in the CS frame,
E866~\cite{bib:e866} observed a small \jpsi\ transverse polarization
($\lambda_{\vartheta}\approx 0.1$) while the HERA-B pattern indicates
longitudinal polarization, of decreasing magnitude with increasing
$p_\mathrm{T}$.  These are not conflicting observations, given the
significantly different \xf\ windows covered: the average \jpsi\
longitudinal momentum, in the centre of mass of the collision system,
is 7 and $-1.4$~GeV/$c$ for E866 and HERA-B, respectively.  Indeed,
Fig.~\ref{fig:all_data_p} shows that the \emph{total} \jpsi\ momentum
(here calculated using average $x_\mathrm{F}$ values) provides a good
scaling between the two fixed-target data sets.
As also shown in Fig.~\ref{fig:all_data_pt}, CDF~\cite{bib:cdf}
reported that, above $p_\mathrm{T}=5$~GeV/$c$, the \jpsi\ polarization
is longitudinal in the H frame, with $\lambda_{\vartheta}$ steadily
decreasing with \pt.  To see how the CDF pattern compares to the
fixed-target data sets, we need to convert the published values to the
CS frame.  We did this translation (using the relations presented
above) assuming that $\lambda_\varphi = 0$ in the CS frame, as
suggested by the HERA-B measurements.  The resulting pattern, seen in
Fig.~\ref{fig:all_data_p}, is perfectly aligned with the HERA-B and
E866 data points.

This smooth overlap of the three data sets
suggests a simple polarization scenario, where the CS frame is taken
to be a good approximation of the natural polarization frame
($\lambda_\varphi = 0$, $\lambda_{\vartheta \varphi} = 0$) and
$\lambda_\vartheta$ is a monotonically increasing function of the
total \jpsi\ momentum.
Before searching a suitable function, we remind that a significant
fraction of the observed \jpsi\ mesons results from $\chi_c$ and
$\psi^\prime$ feed-down decays~\cite{bib:feeddown}: $f_{\rm fd} =
0.33\pm0.05$.
Irrespectively of the possible polarizations of these charmonium
states, it is reasonable to assume that the strong kinematical
smearing induced by the varying decay kinematics reduces the
\emph{observable} polarization of the indirectly produced \jpsi\
mesons to a negligible level.  The feed-down contribution from
$b$-hadron decays can also be neglected: very small at fixed-target
energies and experimentally rejected in the CDF analysis.  Therefore,
the observed polarization should be essentially determined by the
directly produced \jpsi's.
The curve in Fig.~\ref{fig:all_data_p} represents a fit of all the
data points using the simple parametrisation
\begin{equation}
\lambda_\vartheta= (1-f_{\rm fd})\times\left[1-2^{1-(p/p_0)^\kappa}\right]\quad,
\label{eq:lambdath_toymodel}
\end{equation}
where the polarization of the \emph{directly produced} \jpsi's changes
from fully longitudinal at zero momentum to fully transverse at
asymptotically high momentum.  The fit gives $p_0 = 5.0 \pm
0.4$~GeV/$c$ and $\kappa = 0.6 \pm 0.1$, with $\chi^2/{\rm ndf} =
3.6/13$.

\begin{figure}[hb]
\centering
\begin{minipage}[t]{0.48\textwidth}
\resizebox{\textwidth}{!}{%
\includegraphics{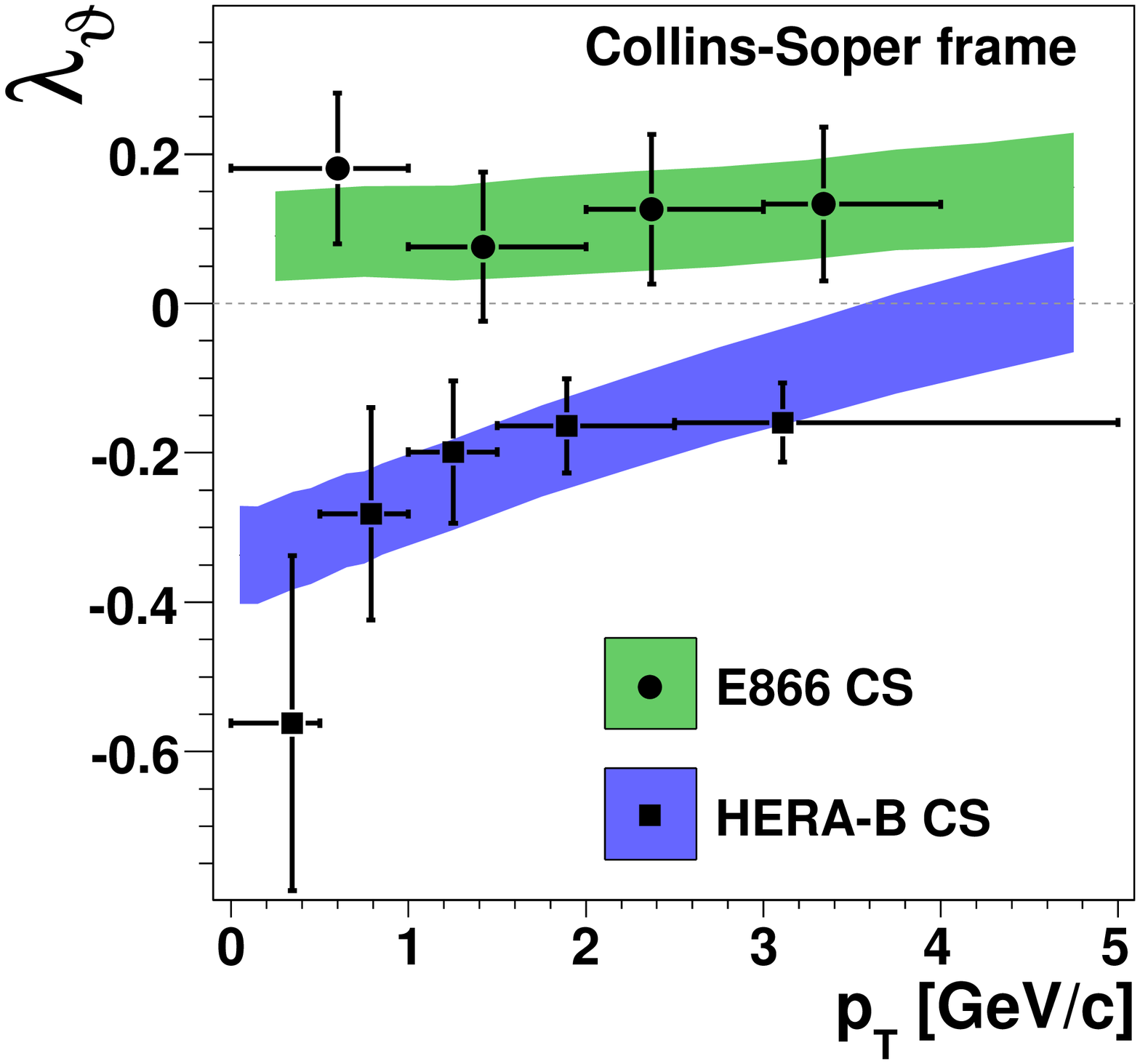}}
\vglue-2mm
\caption{\pt\ dependence of $\lambda_{\vartheta}$ in the CS
  frame, according to Eq.~\ref{eq:lambdath_toymodel} and as reported
  by E866 and HERA-B.}
\label{fig:fit_results_CS}
\end{minipage}
\hfill
\begin{minipage}[t]{0.48\textwidth}
\resizebox{\textwidth}{!}{%
\includegraphics{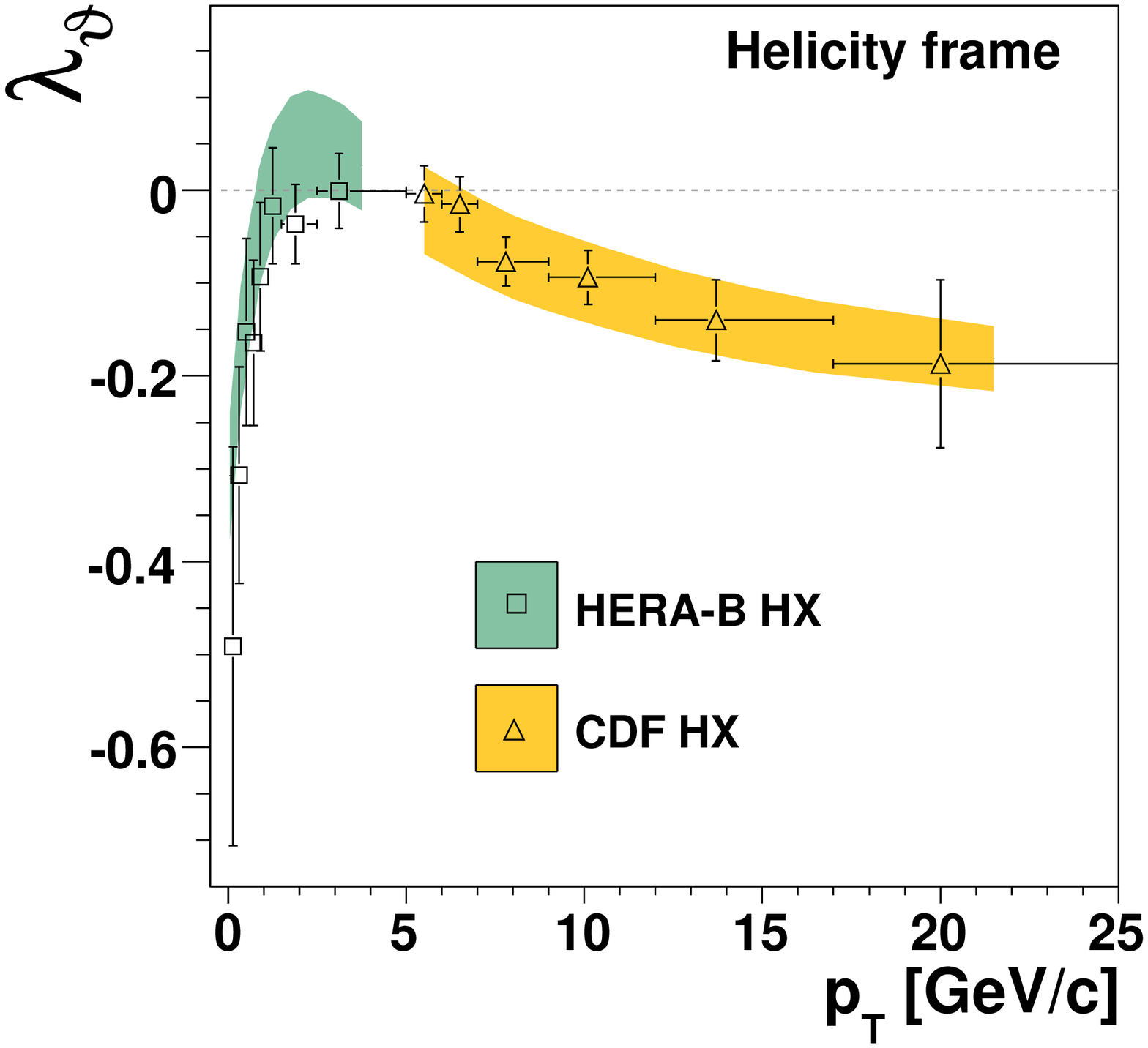}}
\vglue-2mm
\caption{\pt\ dependence of $\lambda_{\vartheta}$ in the H
  frame, as derived from Eq.~\ref{eq:lambdath_toymodel} and as
  reported by HERA-B and CDF.}
\label{fig:fit_results_HX}
\end{minipage}
\end{figure}

\begin{figure}[ht]
\centering
\begin{minipage}[t]{0.48\textwidth}
\resizebox{\textwidth}{!}{%
\includegraphics{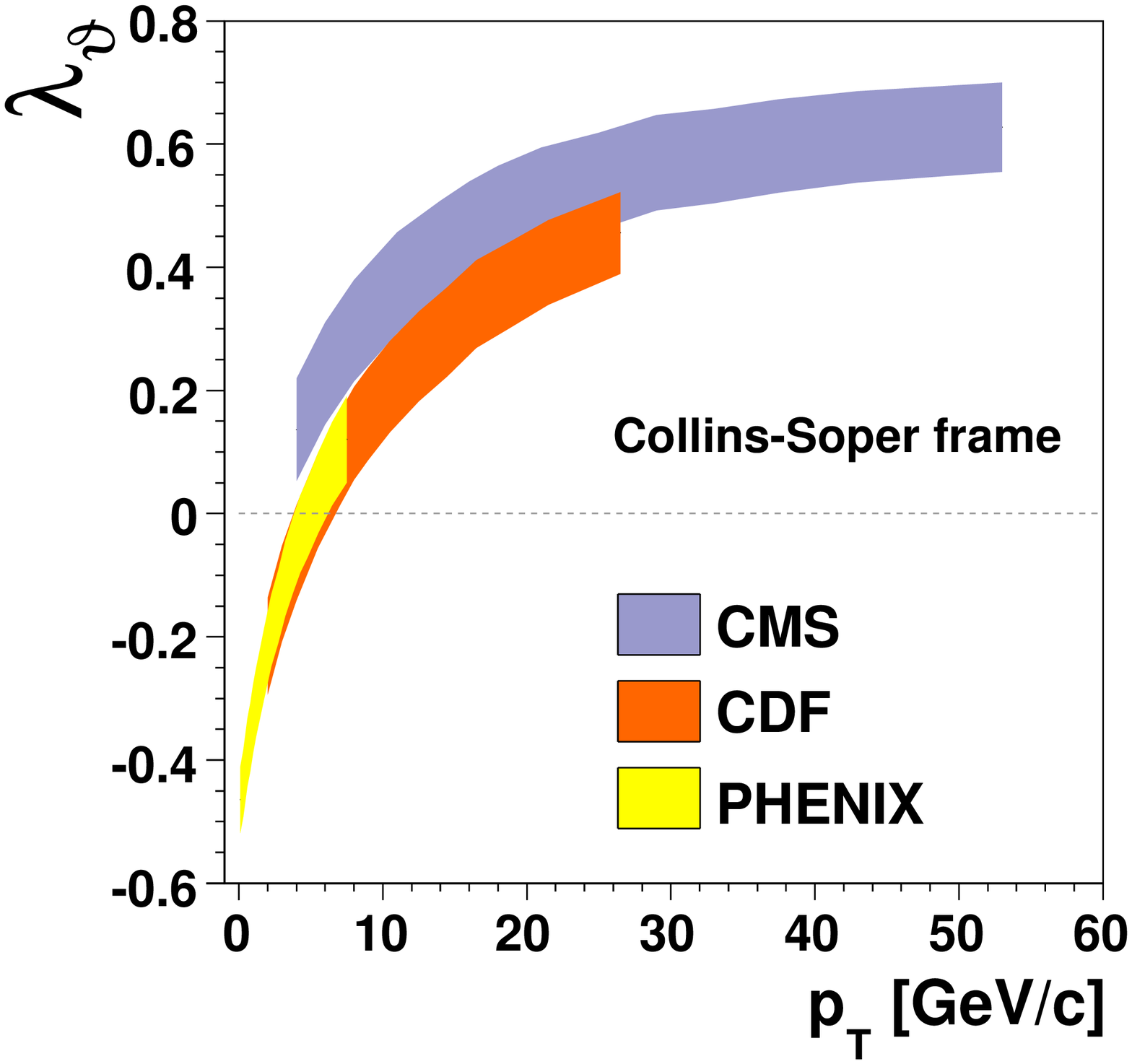}}
\vglue-2mm
\caption{\pt\ dependence of $\lambda_{\vartheta}$ in the CS
  frame, calculated for the energy and rapidity windows of the PHENIX,
  CDF and CMS experiments.}
\label{fig:future_th}
\end{minipage}
\hfill
\begin{minipage}[t]{0.48\textwidth}
\resizebox{\textwidth}{!}{%
\includegraphics{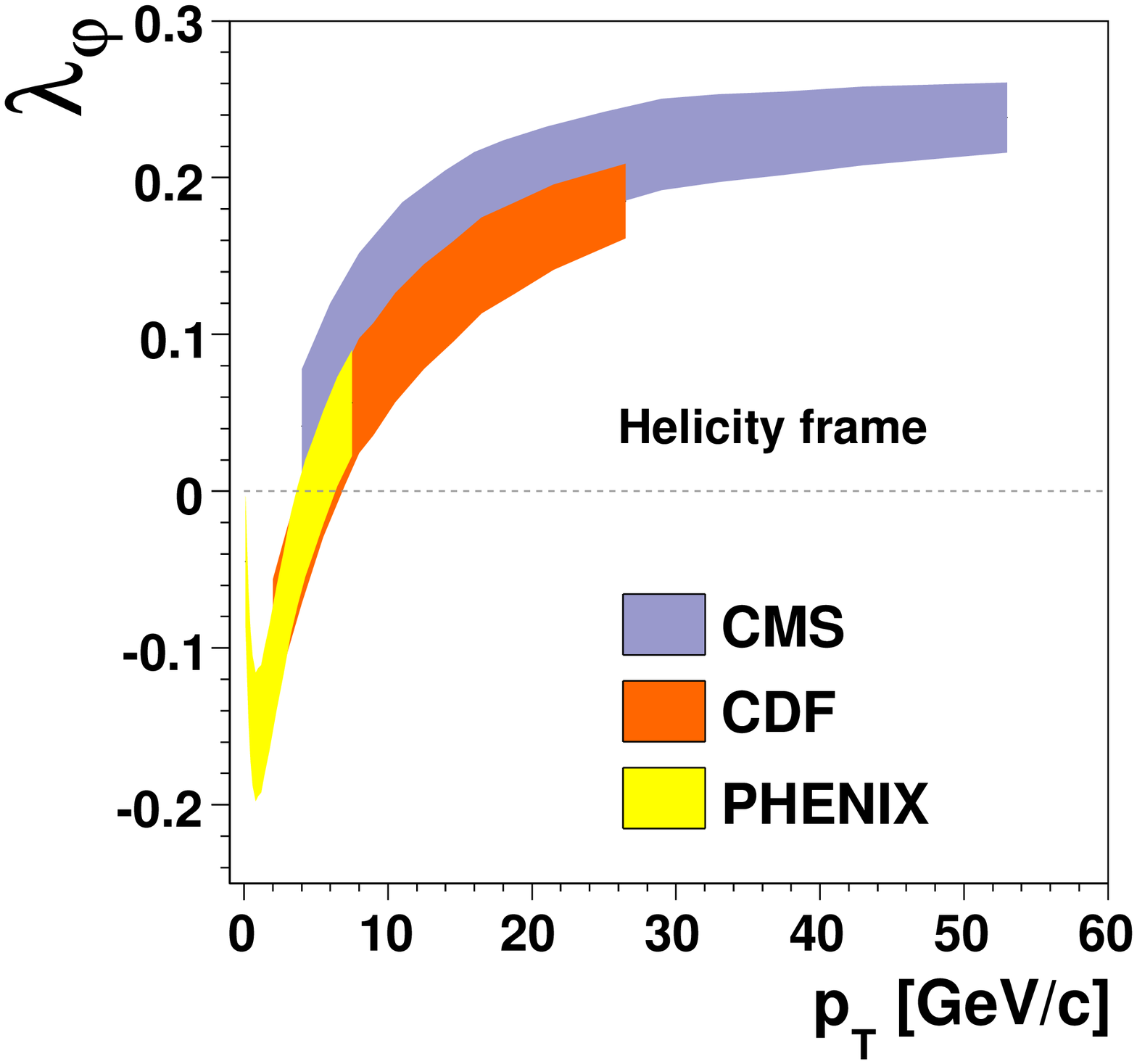}}
\vglue-2mm
\caption{Same as previous figure, but for $\lambda_{\varphi}$ in the
  H frame.}
\label{fig:future_ph}
\end{minipage}
\end{figure}

Our simple parametrisation provides a good description of the existing
data sets, as can be seen in Figs.~\ref{fig:fit_results_CS}
and~\ref{fig:fit_results_HX}, where the widths of the bands correspond
to $\pm 1\sigma$ variations in the two fitted parameters as well as in
the \jpsi\ feed-down fraction.
The derivation of the $\lambda_{\vartheta}$ pattern in the H
frame (needed, in particular, to address the CDF case) incorporates
the ``kinematical smearing'' induced by the decays and the
differential acceptances of the experiments (using a simple Monte
Carlo procedure).
In the narrow rapidity window of CDF, where the maximum
\jpsi\ longitudinal momentum ($\sim$\,4~GeV/$c$) is always smaller
than the minimum $p_\mathrm{T}$ (5~GeV/$c$), the helicity and
Collins-Soper frames are essentially orthogonal to each other.
Therefore, the decrease of $\lambda_{\vartheta}$ with \pt\ seen in the
H frame (Fig.~\ref{fig:fit_results_HX}) is equivalent to an
increase in the CS frame, as shown in
Fig.~\ref{fig:future_th}.

Assuming that the decay distribution has a purely polar anisotropy in
the Collins-Soper frame, with $\lambda_{\vartheta}$ depending on
momentum according to Eq.~\ref{eq:lambdath_toymodel}, CDF should
observe a significant azimuthal anisotropy in the helicity frame, with
a $\lambda_{\varphi}$ pattern (shown in Fig.~\ref{fig:future_ph})
similar in magnitude but of opposite sign with respect to their
$\lambda_{\vartheta}(p_\mathrm{T})$ curve.
By simply repeating the \jpsi\ polarization analysis using the
CS frame and by reporting the azimuthal angular
distribution, CDF can clarify whether the polarization of the \jpsi\
is, also at collider energies, induced along a direction close to the
parton-parton interaction line.

Figures~\ref{fig:future_th} and~\ref{fig:future_ph} also show the
calculated \pt\ dependence of $\lambda_{\vartheta}$, in the
CS frame, and of $\lambda_{\varphi}$, in the H
frame, for the kinematical conditions of the PHENIX
($\sqrt{s}=200$~GeV, $|\eta|<0.35$) and CMS ($\sqrt{s}=14$~TeV,
$|\eta|<2.4$) experiments.
If Eq.~\ref{eq:lambdath_toymodel} remains valid up to LHC
energies, we should see $\lambda_{\vartheta}$ saturating
for \pt\ values higher than those probed by CDF, with a magnitude
determined by the fraction of directly produced \jpsi\ mesons.

We will now summarise our main messages.
1)~To investigate the \jpsi\ polarization and understand its origin,
it is essential to know both the polar and azimuthal distributions,
and their kinematical dependences, in at least two frames.
The Collins-Soper and helicity frames, exactly orthogonal to each
other at mid-rapidity, represent a good minimal set of polarization
frames.
2)~The HERA-B measurements show a pure polar anisotropy in the
CS frame while a mixture of polar and azimuthal
anisotropies is seen in the H frame, indicating that the
\jpsi\ decay angular distribution assumes its simplest shape when
observed with respect to a polar axis that reflects the relative
momentum of the colliding partons rather than the \jpsi\ momentum.
3)~The seemingly contradictory patterns published by E866, HERA-B and CDF
can be consistently reproduced assuming that the polarization (in the
CS frame) of the directly produced \jpsi\ mesons changes
gradually from fully longitudinal at zero momentum to fully transverse
at very high $p$.
4)~This suggests that the longitudinal polarization reported by CDF in
the H frame is, in fact, the reflection of a transverse
polarization (around twice as large) in the CS frame,
increasing with \pt.  Moreover, an azimuthal anisotropy of the decay
distribution should exist in the H frame, with the same
significance as the polar result.
5)~Our polarization scenario predicts that the polar anisotropy of the
prompt \jpsi\ sample will saturate, for \pt\ above $\sim$\,25~GeV/$c$,
at $\lambda_{\vartheta} \approx 0.6$--0.7, a value determined by the
magnitude of the $\psi^\prime$ and $\chi_c$ feed-down contributions,
assumed to be of negligible observable polarization.  This prediction, easily
verifiable at the LHC, can be placed on more robust
grounds once CDF reports the complete angular distribution in the
CS frame or, at least, the azimuthal component in the
H frame.

We acknowledge very stimulating discussions with R.\ Spighi.
P.F.\ and H.W.\ are supported by FCT (Portugal) contracts
SFRH/BPD 42343/2007 and 42138/2007.



\begin{thebibliography}{99}

\bibitem{bib:Theory} E. Braaten, B.A. Kniehl and J. Lee,
  \myprd{62}{2000}{094005};\\ H. Haberzettl and J.P. Lansberg,
  \myprl{100}{2008}{032006}.

\bibitem{bib:cdf} A. Abulencia {\it et~al.} (CDF Coll.),
  \myprl{99}{2007}{132001}.

\bibitem{bib:herab} P. Faccioli for the HERA-B Coll., Int. Workshop on
  Heavy Quarkonium, DESY, Hamburg, October 2007;\\ I. Abt {\it et al.}
  (HERA-B Coll.), Eur. Phys. J. C, in print [arXiv:0901.1015].

\bibitem{bib:e866} T.H. Chang {\it et~al.} (E866 Coll.),
  \myprl{91}{2003}{211801}; T.H. Chang, PhD thesis, NMSU, 1999.

\bibitem{bib:coll_sop} J.C. Collins and D.E. Soper, 
  \myprd{16}{1977}{2219}.

\bibitem{bib:gott_jack} K. Gottfried and J.D. Jackson, Nuovo
    Cim. {\bf 33} (1964) 309.

\bibitem{bib:feeddown} P.~Faccioli {\it et~al.},
  \jhep{10}{2008}{004}.

\end{thebibliography}
\end{document}